\documentclass[twocolumn,eqsecnum,showpacs,preprintnumbers,amsmath,amssymb]{revtex4}
\usepackage{graphicx}
\begin{document}
\preprint{IMAFF-RCA-05-03}
\title{Wiggly cosmic strings accrete dark energy}
\author{Pedro F. Gonz\'{a}lez-D\'{\i}az}\email{p.gonzalezdiaz@imaff.cfmac.csic.es}
\affiliation{Colina de los Chopos, Centro de F\'{\i}sica ``Miguel A.
Catal\'{a}n'', Instituto de Matem\'{a}ticas y F\'{\i}sica Fundamental,\\
Consejo Superior de Investigaciones Cient\'{\i}ficas, Serrano 121,
28006 Madrid (SPAIN)} \author{Jos\'{e} A. Jim\'{e}nez
Madrid}\email{madrid@iaa.es} \affiliation{Colina de los Chopos,
Centro de F\'{\i}sica ``Miguel A. Catal\'{a}n'', Instituto de Matem\'{a}ticas y
F\'{\i}sica Fundamental,\\ Consejo Superior de Investigaciones
Cient\'{\i}ficas, Serrano 121, 28006 Madrid (SPAIN)\\Instituto de
Astrof\'{\i}sica de Andaluc\'{\i}a, Consejo Superior de Investigaciones
Cient\'{\i}ficas, Camino Bajo de Hu\'{e}tor 50, 18008 Granada (SPAIN)}
\date{\today}
\begin{abstract}

This paper deals with a study of the cylindrically symmetric
accretion of dark energy with equation of state $p=w\rho$ onto
wiggly straight cosmic strings. We have obtained that when $w>-1$
the linear energy density in the string core gradually increases
tending to a finite maximum value as time increases for all
considered dark energy models. On the regime where the dominant
energy condition is violated all such models predict a steady
decreasing of the linear energy density of the cosmic strings as
phantom energy is being accreted. The final state of the string
after such an accretion process is a wiggleless defect. It is
argued however that if accreation of phantom energy would proceed
by successive quantum steps then the defect would continue losing
linear energy density until a minimum nonzero value which can be
quite smaller than that corresponding to the unperturbed string.

\end{abstract}


\maketitle

\section{Introduction}

Cosmic strings are known as topological defects that occur in
theories with spontaneous symmetry breaking of a local U(1) gauge
symmetry which were formed during phase transitions in the early
universe. They are trapped infinitely long very thin tubes filled
with a previous false-vacuum phase characterized by a energy density
per unit length $\mu_0=T_0\sim\sigma^2$, with $T_0$ the string
tension and $\sigma$ the symmetry breaking scale, immersed in the a
true-vacuum phase created after the phase transition [1]. Typically,
cosmic strings have been hypothesized as the seeds for ulterior
galaxy formation [2] or as the cosmic sites where primordial
inflation took place [3]. At any event, among the different
theoretical objects that are thought to have populated the universe
at some previous or current periods, which also include black holes
with distinct sizes, Lorentzian wormholes or ringholes, etc., cosmic
strings are the sole objects whose existence has been confirmed in
the laboratory [4]. Therefore, the great interest that cosmic
strings raised when they were first introduced in cosmology has
remained alive all the way up to now. Two general kinds of cosmic
strings have been so far considered, straight strings and string
loops [1]. In the present paper we shall investigate how distinct
forms of dark energy can be accreted by cosmic strings. We shall
restrict ourselves to consider only straight cosmic strings. These
are usually described by a static space-time exterior metric, first
derived by Vilenkin [5]
\begin{equation}
ds^2 = -dt^2+dr^2+dz^2+(1-8G\mu_0)r^2 d\phi^2 .
\end{equation}
By defining a new cylindrical angular coordinate $\phi
'=(1-8G\mu_0)\phi$, it can immediately be seen that this metric
corresponds to a flat spacetime with a conical singularity that is
associated with a deficit angle given by $\Delta=8\pi G\mu_0$.

However, an incoming or outgoing energy flux due to dark energy
accretion is no longer strictly possible for an exterior locally
flat metric like that of a motionless straight string having no
wiggles [1]. In fact, a motionless string with no wiggles [1]
cannot accrete anything that is motionless and homogeneous around
it -in particular it could not accrete dark energy. Thus, if we
want to consider accretion of dark energy onto cosmic strings we
need these cosmic string to be perturbed by wiggles. In that case
the exterior string metric can no longer be given by the locally
flat line element (1.1), as wiggle-induced variations of the
string mass per unit length and tension would convert these
quantities into space-time dependent functions, $\mu$ and $T$,
with the state equation $\mu T=\mu_0^2$ and $\mu>T$, whose values
can initially be considered to be very similar to each other and
therefore also very similar to their unperturbed counterparts in
the linear approximation [1]. The linearized wiggly string metric
reads [1]
\begin{eqnarray}
&&ds^2=-\left[1+4G\left(\mu-T\right)\ln\frac{r}{r_0}\right]dt^2+dr^2\nonumber\\
&&+\left[1-4G\left(\mu-T\right)\ln\frac{r}{r_0}\right]dz^2\nonumber\\
&&+\left[1-4G\left(\mu+T\right)\right]r^2 d\theta^2 ,
\end{eqnarray}
which, contrary to metric (1.1), produces a non-vanishing Newtonian
potential. In this case the deficit angle is given by $4\pi
G(\mu+T)$.

Nowadays cosmology, on the other hand, relies mainly on the idea
that the total energy of the current universe and possibly that of
the early universe (that is the two cosmic periods known to show
accelerating expansion) is dominated by some form of the so-called
dark energy [6]. It is therefore of interest to investigate the
effects that dark energy may cause in cosmic strings. Following
the recent studies performed on black holes [7,8], one can
actually suppose that dark energy can also be accreted onto a
cosmic string, inducing some variation in its energy density per
unit length $\mu$. This work aims at considering the effects that
the accretion of dark energy may have in the fate of wiggly
straight cosmic string in an accelerating universe. We shall
represent dark energy as a perfect fluid characterized by a
negative parameter $-1/3>w=p/\rho$ (with $p$ the pressure and
$\rho$ the energy density) filling a Friedmann-Robertson-Walker
universe whose scale factor is given by [9]
\begin{equation}
a(t)=a_0\left(1+\frac{3}{2}(1+w)C^{1/2}(t-t_0)\right)^{2/[3(1+w)]}
,
\end{equation}
where $C=8\pi G\rho_0/3$ and we have taken for the energy density
$\rho=\rho_0 a^{-3(1+w)}$, with $\rho_0$ an integration constant,
if we adopt a general quintessence model. It can be readily seen
that, whereas the universe enters a steady regime of accelerating
expansion which keeps it being finite all the way up to an
infinite time if $w>-1$, in the case that $w<-1$ (a case at which
the dark energy is called phantom energy [10]) the universe would
expand along super-accelerated patterns that drive it to a
singularity at a finite time in the future at which everything
-even the elementary particles - loses any independent, local
behavior by its own to be ripped apart under the sole influence of
the global phantom cosmological law. This singularity has been
dubbed the big rip [11] and takes place at a time
\begin{equation}
t_* =t_0 +\frac{2}{3(|w|-1)C^{1/2}} .
\end{equation}

Such a rather weird behavior takes also place when the other main
contender model for dark energy, that is to say the K-essence model
[12], is assumed to dominate. In fact, if $w<-1$ we obtain in this
case [13]
\begin{equation}
a(t)\propto (t-t_*)^{-2\beta/[3(1-\beta)]} ,\;\;\; 0<\beta<1 ,
\end{equation}
where $t_*$ again represents the time for the big rip which is an
arbitrary parameter in this case.

Of particular interest is the scenario in which we consider that the
dark energy is given in terms of a generalized Chaplygin gas having
an equation of state $p=-A\rho^{-\alpha}$, with $A>0$ and $\alpha$ a
parameter [14]. In this case the cosmic time $t$ relates to the
scale factor by the more complicated expression [15]
\begin{eqnarray}
&&t-t_0=\frac{2\left[1+\frac{B}{A}a^{-3(1+
\alpha)}\right]^{\frac{1+2\alpha}{2(1+
\alpha)}}}{3(1+2\alpha)\sqrt{CA^{1/(1+
\alpha)}/\rho_0}}\times\nonumber\\
&&F\left(1,\frac{1+2\alpha}{2(1+\alpha)};\frac{3+
4\alpha}{2(1+\alpha)}; 1+\frac{B}{A}a^{-3(1+\alpha)}\right) ,
\end{eqnarray}
with $F$ a hypergeometric function and
$B=(\rho^{1+\alpha}-A)a_{0}^{3(1+\alpha)}$. It can be seen that even
in the phantom energy regime a generalized Chaplygin gas does not
lead to any big rip singularity in the future [15,16], but it always
drives a steady regular accelerating expansion for the universe.

In this paper we use a formalism which is able to encompass the
accretion of dark energy described by any of the above models onto
wiggly straight cosmic strings. We obtain that as quintessence or
K-essence dark energy is accreted onto a perturbed straight cosmic
string the energy density per unit length of this string either
progressively increases up to a constant finite value if $w>-1$,
or steadily decreases down to the unperturbed value first and
might then enter a region where quantum accretion makes it reach a
minimum value, quite before the occurrence of the big rip
singularity if $w<-1$. The behavior of the strings when they
accrete Chaplygin gas is similar: their energy density per unit
length also progressively either increases or decreases toward a
extremal value, depending on whether the dominant energy condition
is satisfied or violated.

The paper can be outlined as follows. In Sec. II we present the
general formalism for the accretion of dark energy onto straight
wiggly cosmic strings and obtain a general rate equation for the
string core energy density per unit length, $\mu$, in terms of the
internal dark energy, and apply such a formalism to quintessence
and K-essence cosmological fields, so as to the generalized
Chaplygin gas model. Approximate expressions of $\mu$ as a
function of time for the first two dark energy models are also
derived, both for $w=p/\rho>-1$ and $w=p/\rho<-1$, analyzing the
corresponding evolution of the cosmic strings. Finally we conclude
and add some further comments in Sec. III.

\section{Dark energy accretion onto wiggly straight cosmic strings}

We shall consider next how the general accretion theory can be
applied to the case in which dark energy is accreted onto wiggly
cosmic strings. We shall generally follow the procedure put
forward by Babichev, Dokuchaev and Eroschenko [7] for the case of
Schwarzschild black holes, generalizing it to the case of straight
wiggly cosmic strings. Thus, we start by integrating the
energy-momentum conservation law by using the exterior metric
(1.2). Although for metric (1.1) there are only two non-vanishing
components of the Christoffel symbols,
$\Gamma^r_{\theta\theta}=-\left(1-8G\mu\right)r$ and
$\Gamma^\theta_{r\theta}=1/r$, when the string is perturbed with
wiggles there will be twenty one generally non-vanishing
components of the Christoffel symbols which make the calculation
to follow more complicated. For a cylindrical symmetry we then
have from the time-component of the conservation law of the
energy-momentum tensor, $T_{\mu;\nu}^{\nu}=0$,
\begin{equation}
\sqrt{\mu}ru\sqrt{1-h_{00}}\sqrt{1-b}(1+h_{00})\sqrt{u^2-1}(p+\rho)=C
,
\end{equation}
where
\begin{equation}
h_{00}=4G(\mu-T)\ln\left(r/r_0\right)
\end{equation}
\begin{equation}
b=4G(\mu+T),
\end{equation}
with $r_0$ and $C$ integration constants and $u=dr/ds$.

After integrating the conservation law for the energy-momentum
tensor projected onto the four-velocity,
$u_{\mu}T_{;\nu}^{\mu\nu}=0$, we also obtain
\begin{equation}
ur\sqrt{\mu(1-h^2_{00})(1-b)}e^{\int_{\rho_{\infty}}^{\rho}\frac{d\rho}{p+\rho}}=A,
\end{equation}
where we have taken into account that $u$ should be positive for
incoming energy flux in this case, and $A$ is a positive constant.
From Eqs. (2.1) and (2.4) we can then get
\begin{equation}
\sqrt{(u^2-1)(1+h_{00})}(p+\rho)e^{-\int_{\rho_{\infty}}^{\rho}\frac{d\rho}{p+\rho}}=C_2,
\end{equation}
in which the constant $C_2$ can be expressed as
$C_2=C/A=\hat{A}\left[\rho_{\infty}+p(\rho_{\infty})\right]$, with
$\hat{A}>0$ a constant, for the cylindrical symmetry used.

By integrating now the momentum density $T^r_0$ over the circular
length element of the cylinder we can obtain the rate of change of
the energy per unit length of the wiggly cosmic string, so that
\begin{equation}
\dot{\mu}=-\int_0^{2\pi}rT^r_0
d\phi=\int_0^{2\pi}r(p+\rho)(1+h_{00})\frac{dt}{ds}\frac{dr}{ds}d\phi
.
\end{equation}
Using then the property
$\sqrt{1+h_{00}}dt=\sqrt{\frac{dr^2}{ds^2}-1}ds$ stemming from the
cylindrical symmetry being used and Eqs. (2.5) and (2.6), we
finally derive the relevant rate equation for the energy density
of a wiggly cosmic string
\begin{equation}
\dot{\mu}=\frac{2\pi\bar{A}\left[\rho_{\infty}+
p(\rho_{\infty})\right]}{\sqrt{\mu(1-b)(1-h_{00}^2)}} ,
\end{equation}
with $\bar{A}=A\hat{A}>0$ a constant. Therefore, one has the
following integral expression that governs the evolution of the
wiggled mass per unit length of the cosmic string
\begin{equation}
\int_{\mu_i}^{\mu}\sqrt{\mu(1-b)(1-h_{00}^2)}d\mu=
2\pi\bar{A}\int_{t_0}^{t}\left[\rho_{\infty}+p\left(\rho_{\infty}\right)\right]dt
.
\end{equation}
It is worth noticing that the above expressions restrict by
themselves the interval along which the quantity $\mu$ is allowed to
vary on its real values. In fact, one can derive the two conditions
\begin{equation}
\mu_0 < \frac{1}{8G}
\end{equation}
\begin{equation}
\frac{1-\sqrt{1-64G^2\mu_0^2}}{8G}<\mu<\frac{1+\sqrt{1-64G^2\mu_0^2}}{8G}
.
\end{equation}
Condition (2.9) expresses nothing but the impossibility for an
supermassive wiggleless cosmic string to reach a linear energy
density larger than nearly $1/G$. Even though the concepts of radius
and mass per unit length for a source like the string core are not
unambiguously defined [17], specially in the presence of an
interacting dark energy fluid, at the extreme supermassive case
$\mu=1/8G$ one would expect the string to no longer exist because it
then corresponded to the situation where all the exterior broken
phase is collapsed into the core, leaving a pure false-vacuum phase
in which the picture of a cosmic string with a core region of
trapped is lost [18]. When the string is wiggled then condition
(2.9) reflects into condition (2.10) by which it is seen that a
wiggly cosmic string cannot exceed a given maximum value or be less
than a given minimum nonzero value. If a cosmic string has the
extreme supermassive linear mass density, then it cannot be wiggled
nor accrete any kind of dark energy.

Now, the integration in the left-hand-side of Eq. (2.8) appears to
be very difficult to perform and, in fact, we have been unable to
obtain an integrated expression from it in closed form.
Nevertheless, in the physically relevant cases that $\mu$ is very
close to $\mu_0$ and/or $r$ is very close to $r_0$, that term can
be integrated to approximately give
\begin{eqnarray}
&&\int_{\mu_i}^{\mu}d\mu\sqrt{\mu(1-b)(1-h_{00}^2)}\simeq I(\mu)=
\nonumber\\
&&\left[\frac{8G\mu-1}{16G}\sqrt{-4G\mu^2+\mu-4G\mu_0^2}\right.\nonumber\\
&&\left.\left.+\frac{64G^2\mu_0^2-1}{64G^{3/2}}\arcsin\left(\frac{1-
8G\mu}{\sqrt{1-64G^2\mu_0^2}}\right)\right]\right|_{\mu_i}^{\mu}
 .
\end{eqnarray}

The integration of the right-hand-side of Eq. (2.8) will be
performed in what follows for the distinct dark energy models
considered in the Introduction.

\subsection{Quintessence and K-Essence}
Starting with the equation of state $p=w\rho$, where $w$ is
assumed constant, we can use the conservation of cosmic energy to
finally derive
\begin{equation}
\rho=\rho_0\left(\frac{a_0}{a}\right)^{3(1+w)} ,
\end{equation}
with $\rho_0$ and $a_0$ constants. Hence
\begin{eqnarray}
&&2\pi\bar{A}\int_{t_0}^{t}\left[\rho_{\infty}+p\left(\rho_{\infty}\right)\right]dt
=\nonumber\\ &&2\pi\bar{A}(1+w)\rho_0 a_0^{3(1+w)}\int_{t_0}^t dt
a^{-3(1+w)}.
\end{eqnarray}
We then have for the scale factor (1.3) corresponding to a general
flat quintessence universe
\begin{equation}
t=t_0+\frac{I(\mu)}{(1+w)\left(2\pi\bar{A}\rho_0-\frac{3}{2}C^{1/2}I(\mu)\right)}
,
\end{equation}
where $I(\mu)$ is defined in Eq. (2.11). This is a parametric
equation from which one can obtain how the energy per unit length
of a wiggled cosmic string evolves in the accelerating universe.
Thus, if $w>-1$ we see that the string energy in the core will
progressively increases from its initial value $\mu_i$, tending to
the maximum value
\[\mu_{\rm max}=\frac{1+\sqrt{1-64G^2\mu_0^2}}{8G}.\]
The larger $w$ the shorter the time required by the accretion
process to make the string to reach $\mu_{\rm max}$. If $w<-1$,
i.e. if we are in the phantom regime, then the linear energy
density in the string core will rapidly decreases from its initial
value down to recover its unperturbed value at $\mu_0$. The
smaller $w$ the shorter the time taken by the system to reach the
value $\mu_0$. As the string is approaching that value the
gravitational potential should be getting on smaller and smaller
values to finally vanish at $\mu_0$, so that the classical
accretion process will stop at that point. Such a behavior is also
checked to occur in the case that phantom K-energy is accreted.

An interesting question is however posed in the two considered
kinds of phantom energy. Even though the classical, continuous
accretion process must only proceed down to $\mu_0$, if we assumed
that phantom energy accretion would proceed by discrete steps,
then the limit at $\mu_0$ should be overtaken and the linear
energy density of the string core would continue decreasing below
$\mu_0$ as the phantom energy was being accreted. We would reach
in this way a regime where $T>\mu$ which would end when $\mu$
reached the minimum value
\[\mu_{\rm min}=\frac{1-\sqrt{1-64G^2\mu_0^2}}{8G} ,\]
which would never vanish provided $\mu_0>0$. The spacetime metric of
the cosmic string given by Eq. (1.2) would then exchange the values
between the $tt$ and $zz$ components, as in this case $\mu<T$.

\subsection{Generalized Chaplygin gas}
We shall derive now the expression for the rate $\dot{\mu}$ in the
case of a generalized Chaplygin gas. We start with the expression
for the energy density
\begin{equation}
\rho=\left(A_{ch} +\frac{B}{a^{3(1+\alpha)}}\right)^{1/(1+\alpha)} ,
\end{equation}
which has been obtained by integrating the cosmic conservation law
for the case of the equation of state of a generalized Chaplygin
gas, that is $p=-A_{ch}/\rho^{\alpha}$. Now, from the Friedmann
equation we can get
\begin{equation}
\dot{a}=\sqrt{\frac{8\pi
G}{3}}a(t)\left(A_{ch}+\frac{B}{a^{3(1+\alpha)}}\right)^{1/[2(1+\alpha)]}
.
\end{equation}
Hence, from Eq. (2.11) it can be obtained
\begin{eqnarray}
&&I(\mu)=B\bar{A}\sqrt{\frac{3\pi}{2G}}\int_{a_0}^{a}
\frac{\frac{1}{a^{3(1+\alpha)}}}{a\left(A_{ch}+
\frac{B}{a^{3(1+\alpha)}}\right)^{(2\alpha+1)/[2(1+\alpha)]}}da
\nonumber\\
&&=-\bar{A}\sqrt{\frac{2\pi}{3G}}\left[\left(A_{ch}+\frac{B}{a^{3(1+
\alpha)}}\right)^{1/[2(1+\alpha)]}- \sqrt{\rho_0}\right].
\end{eqnarray}
It follows
\begin{equation}
a^{3(1+\alpha)}=\frac{B}{\left(\sqrt{\rho_0}-\sqrt{\frac{3G}{2\pi
\bar{A}^2}}I(\mu)\right)^{2(1+\alpha)}-A_{ch}} .
\end{equation}
Again in this case the setting of a constant $B>0$ implies a
progressive increase of $\mu$ with $a$ up to a maximum given by
$\mu_{\rm max}$, and the assumption of a constant $B<0$ (phantom)
leads to a decrease of $\mu$ with $a$ down to $\mu_0$ classically or
to $\mu_{\rm min}$ if the Chaplygin phantom energy is supposed to be
accreted in discrete steps.

\section{Conclusions and further comments}

While cosmic strings have a long tradition and incidence in
theoretical cosmology, the introduction of cosmic dark energy has
taken place quite more recently though not with less incidence or
surprise. Perhaps therefore their potential mutual relations and
interactions have not been so far considered. This paper is a
first step in the task of studying the effects that the presence
of dark energy may have in the fate of cosmic string in an
accelerating universe. We have restricted ourselves here to just
looking at an approximate model describing how straight wiggly
cosmic strings accrete dark energy during the accelerating
expansion of the universe, leaving for future publications the
accurate treatment for both wiggly straight strings and the
similar accretion onto circular strings, so as the kinematic
effects that the acceleration of the universe may have on the
shape and size of any cosmic strings. A generalized description
has first been thus built up and then adapted to the case of the
cylindrically symmetric accretion of dark energy onto straight
cosmic strings. That description is based on the integration of
the conservation laws for the energy-momentum tensor and its
projection on four-velocity using the exterior geometry of a
wiggly cosmic string. We have considered the dark energy accretion
onto straight cosmic strings using several scalar field models for
the cosmological vacuum, namely quintessence and K-essence field
models with equation of state $p=w\rho$, and a generalized model
of Chaplygin gas with the unusual equation of state
$p=-A/\rho^{\alpha}$. An rate equation for the energy density per
unit length of the strings has been in this way derived and
finally integrated for each of these dark energy models. This
ultimately leads to the prediction that, whereas when the energy
density of the cosmic vacuum decreases with time the linear energy
density of the straight strings progressively increases as the
universe grows bigger for all dark energy models, if the energy
density of the universe grows with expansion, inducing a universal
violation of the dominant energy condition, the stringy energy
density steadily decreases. That energy density dropping makes the
strings to eventually become free of wiggles to get thereafter on
a quantum accreating regime where the string energy density
reaches finally a minimum nonzero value, before the occurrence of
any future big rip singularities.

It appears that the current value of the parameter $w$ in the
equation of state of the universe may be less than -1. So, one
could be tempted to think that the above evolution of cosmic
strings leading eventually to the formation of exotic topological
defects with negative-wiggles perturbations would be inescapable.
However, having now $w<-1$ (provided this turns out to be
definitively the case most favoured by observations) does not
guarantee at all that the phantom regime will endure in the
future. In fact, most general descriptions of quintessence field
are based on tracking models where the parameter $w$ is time
dependent [19] and, therefore, it could well be that what is now
less than -1 would later turn out to be greater than -1, so making
the cosmic string evolution predicted by our constant-$w$ models
inapplicable in the far future. Nevertheless, the initial string
evolution implied by our phantom models looks as being probable.
That behaviour by itself would still be important enough for a
variety of subjects. But even such a behaviour would not be
guaranteed as phantom fields are characterized by Lagrangians
containing negative kinetic terms which have very weird properties
and lead to unwanted instabilities [20] making the whole phantom
scenario problematic.

\acknowledgements

\noindent We thank Professors J.A.S. Lima and E. Babichev for
useful explanations, discussions and correspondence. We also
acknowledge A. Ferrera and M. Rodr\'{\i}guez for constructive
discussions and criticisms. This work was supported by DGICYT
under Research Projects BMF2002-03758 and BFM2002-00778. JAJM
wants to acknowledge IMAFF for kind hospitality.

\pagebreak


\begin{references}
\bibitem {1} T.W.B. Kibble, J. Phys. A9, 1387 (1976); A. Vilenkin,
Phys. Rep. 121, 236 (1985); A. Vilenkin and E.P.S. Shellard, {\it
Cosmic Strings and Other Topological Defects} (Cambridge
University Press, Cambridge, UK, 1994).
\bibitem {2} R.H. Brandenberger, J. Phys. G: Nucl. Part. Phys. 15,
1 (1989); D.N. Vollick, Phys. Rev. D45, 1884 (1992).
\bibitem {3} A.D. Linde and D.A. Linde, Phys. Rev. D50, 2456 (1994);
A. Vilenkin, Phys. Rev. Lett. 72, 3137 (1994).
\bibitem {4} T.W.B. Kibble, {\it Testing Cosmological Defect
Formation in the Laboratory}, cond-mat/0111082, Proceedings of the
Second European Conference on Vortex Matter in Superconductors,
Crete (2001), to appear.
\bibitem {5} A. Vilenkin, Phys. Rev. Lett. 46, 1169 (1981); Phys.
Rev. D24, 2082 (1981).
\bibitem {6} T. Padmanabhan, {\it Dark Energy: The Cosmological
Challenge of the Millennium}, astro-ph/0411044, Current Science (to
appear).
\bibitem {7} E. Babichev, V. Dokuchaev and Yu. Eroschenko, Phys.
Rev. Lett. 93, 021102 (2004); P.F. Gonz\'{a}lez-D\'{\i}az and C.L.
Sig\"{u}enza, Phys. Lett. B589, 78 (2004).
\bibitem {8} P.F. Gonz\'{a}lez-D\'{\i}az and C.L. Sig\"{u}enza, Nucl. Phys.
B697, 363 (2004).
\bibitem {9} C. Wetterich, Nucl. Phys. B302, 668 (1988); J.C. Jackson
and M. Dodgson, Mon. Not. R. Astron. Soc. 297, 923 (1998); J.C.
Jackson, Mon. Not. R. Astron. Soc. 296, 619 (1998); R.R. Caldwell,
R. Dave and P.J. Steinhardt, Phys. Rev. Lett. 80, 1582 (1998); L.
Wang and P.J. Steinhardt, Astrophys. J. 508, 483 (1998); R.R.
Caldwell and P.J. Steinhardt, Phys. Rev. D57, 6057(1998); G. Huey,
L. Wang, R. Dave, R.R. Caldwell and P.J. Steinhardt, Phys. Rev.
D59, 063005 (1999); P.F. Gonz\'{a}lez-D\'{\i}az, Phys. Rev. D62, 023513
(2000); J.D. Barrow, Phys. Lett. B180, 335 (1986); B235, 40
(1990); D. Bazcia and J. Jackiw, Ann. Phys. 270, 246 (1998).
\bibitem {10} R.R. Caldwell, Phys. Lett. B545, 23 (2002)
\bibitem {11} R.R. Caldwell, M. Kamionkowski and N.N. Weinberg,
Phys. Rev. Lett. 91, 071301 (2003); S. Nojiri and S.D. Odintsov,
Phys. Lett. B562, 147 (2003); L.P. Chimento and R. Lazkoz, Phys.
Rev. Lett. 91, 211301 (2003); J.D. Barrow, Class. Quant. Grav. 21,
L79 (2004); S. Nesseris and L. Perivolaropoulos, {\it The Fate of
Bound Systems in Phantom and Quintessence Cosmologies},
astro-ph/0410309 .
\bibitem {12} C. Armend\'{a}riz-Pic\'{o}n, T. Damour and V. Mukhamov,
Phys. Lett. B458, 209 (1999); J. Garriga and V. Mukhamov, Phys.
Lett. B458, 219 (1999).
\bibitem {13} P.F. Gonz\'{a}lez-D\'{\i}az, Phys. Lett. B586, 1 (2004).
\bibitem {14} A. Kamenshchik, U. Moschella and V. Pasquier, Phys.
Lett. B511, 265 (2001); N. Bilic, G.B. Tupper and R. Viollier,
Phys. Lett. B535, 17 (2001).
\bibitem {15} M. Bouhmadi and J.A. Jim\'{e}nez Madrid, JCAP 0505 005 (2005).
\bibitem {16} P.F. Gonz\'{a}lez-D\'{\i}az, Phys. Rev. D68, 021303  (2003).
\bibitem {17} See for example, G.W. Gibbons, S.W. Hawking and T.
Vachaspati, {\it The Formation and Evolution of Cosmic Strings}
(Cambridge University Press, Cambridge, UK, 1990).
\bibitem {18} P. Laguna and D. Garfinkle, Phys. Rev. D40, 1011 (1989);
M.E. Ortiz, Phys. Rev. D43, 2521 (1991).
\bibitem {19} I. Zlatev, L. Wang and P.J. Steinhardt, Phys. Rev.
Lett. 82, 896 (1999); P.J. Steinhardt, L. Wang and I. Zlatev,
Phys. Rev. D59, 123504 (1999); P. Brax, J. Martin and A. Riazuelo,
Phys. Rev. D62, 103505 (2000).
\bibitem {20} S.M. Carroll, M. Hoffman and M. Trodden, Phys. Rev.
D68, 023509 (2003).


\end{references}
\end{document}